\documentclass{optica-article}

\journal{opticajournal} 

\articletype{Research Article}

\usepackage{lineno}
\usepackage{threeparttable}

\begin{document}

\title{Efficient finite element modeling of photonic modal analysis augmented by combined symmetry}

\author{JINGWEI WANG,\authormark{1} LIDA LIU,\authormark{1} YUHAO JING,\authormark{1} ZHONGFEI XIONG,\authormark{1,*} and YUNTIAN CHEN\authormark{1,2,3,**}}

\address{\authormark{1}School of Optical and Electronic Information, Huazhong University of Science and Technology, Wuhan 430074, China\\
\authormark{2}Wuhan National Laboratory of Optoelectronics, Huazhong University of Science and Technology, Wuhan 430074, China\\
\authormark{3}Optics Valley Laboratory, Wuhan 430074, China}

\email{\authormark{*}xiongzf94@outlook.com\\
\authormark{**}yuntian@hust.edu.cn
}


\begin{abstract*} 
In this work, we present an efficient numerical implementation of the finite element method for modal analysis that leverages various symmetry operations, including spatial symmetry in point groups and space-time symmetry in pseudo-Hermiticity systems. We provide a formal and rigorous treatment, specifically deriving the boundary constraint conditions corresponding to symmetry constraints. Without loss of generality, we illustrate our approach via computing the modes of optical waveguides with complex cross-sections, accompanied with performance benchmark against the standard finite element method. The obtained results demonstrate excellent agreement between our method and standard FEM with significantly improved computational efficiency. Specifically, the calculation speed increased by a factor of $23$ in the hollow-core fiber. Furthermore, our method directly classifies and computes the modes based on symmetry, facilitating the modal analysis of complex waveguides.
\end{abstract*}

\section{Introduction}
In photonics, symmetry serves as a fundamental concept that simplifies our understanding of light-matter interactions in complex optical medium or structures. Symmetry principle not only facilitates the discovery and predication of novel optical phenomena, including symmetry-protected bound states in the continuum\cite{jin_topologically_2019,liu_circularly_2019,dong_nanoscale_2022,wang_continuum_2023}, optical Weyl points\cite{chang_multiple_2017,yang_ideal_2018,pan_real_2023}, topological edge states\cite{khanikaev_photonic_2013,song_detailed_2024} and enhanced optical nonlinearities\cite{konotop_nonlinear_2016,zhang_symmetry-breaking-induced_2019}, but also inspires the design of photonic devices. These devices, including optical fibers\cite{zhao_polarization-maintaining_2017,liu_symmetrical_2018,gao_design_2019,murphy_azimuthal_2023}, metamaterials\cite{cong_symmetry-protected_2019,droulias_chiral_2019}, lasers\cite{feng_single-mode_2014,kodigala_lasing_2017,hokmabadi_supersymmetric_2019}, and resonators\cite{chang_paritytime_2014,chen_parity-time-symmetric_2018,mortensen_fluctuations_2018,kim_parity-time_2024}, utilize symmetry principle to deliver exceptional performance. Despite its significance, symmetry principle has not been systematically studied in the numerical modeling of photonic devices, especially in full-wave methods such as the finite element method (FEM) and finite difference method (FDM).

Recent work on using symmetry principle in numerical modeling primarily involves the use of simple symmetry to reduce computational load. For instance, mirror symmetric photonic devices can be modeled by a fraction (half or a quarter) of the device by employing a perfect electric conductor boundary condition (PEC) or a perfect magnetic conductor boundary condition (PMC) to truncate the computational domain\cite{nyakas_full-vectorial_2007,zamani_aghaie_birefringence_2010,otin_finite_2011,andonegui_finite_2013}. In the numerical modeling of periodic devices\cite{hiett_application_2002,nicolet_modelling_2004,tavallaee_finite-element_2008,parisi_complex_2012,bagheriasl_bloch_2019}, such as photonic crystals and metamaterials, the Bloch boundary condition derived from discrete translational symmetry facilitates modeling solely on the unit cell. However, many photonic devices exhibit multiple spatial symmetries, conforming simultaneously to various mirror and rotational symmetries. Taking the hollow-core fiber\cite{gao_hollow-core_2018,roth_strong_2018,kelly_gas-induced_2021,mulvad_kilowatt-average-power_2022,murphy_azimuthal_2023} as an example, which are widely studied due to its superior low-loss performance, such waveguide supports complex modes and poses challenges in analysis and design owing to its high $C_{N,v}$ point group symmetry and large cross-sectional size. Moreover, pseudo-Hermiticity systems\cite{mostafazadeh_pseudo-hermiticity_2002,borgnia_non-hermitian_2020,stalhammar_classification_2021} with space-time symmetries, such as PT-symmetry, predominantly depend on constructing specific forms of Hamiltonian to account for the relation between symmetry-protected degenerate modes, wherein the explicit Hamiltonian in optical system is difficult to extract. Identifying all symmetries of photonic devices and combining these symmetries into modal analysis can significantly improve the efficiency of analysis and reveal symmetry characteristics of the modes prior to the completion of the computation. In this regard, exploring how to apply combined symmetries in numerical modeling is relevant and necessary, though remains elusive.

This work introduces a systematic method for finite element modeling of modal analysis by combining various symmetries for the first time. The method utilizes both the spatial symmetry of point groups and the space-time symmetry of pseudo-Hermiticity systems to simplify the computational complexity. The essence is to decompose the original modal problem into several decoupled sub-tasks, each corresponding to modes of different symmetry types, and subsequently restore the original one based on those solved sub-tasks. The symmetry group guarantees the complete closure between the sub-tasks and the original problem. Compared to the original problem, each sub-task has fewer degrees of freedom (DOFs), thereby significantly reducing total computational time. In the first numerical example, the calculation speed increased by a factor of $4$, while in the second numerical example, the calculation speed increased by a factor of $23$. Comparative numerical examples against standard FEM validate the effectiveness and efficiency of our method. This method facilitates efficient and swift modal analysis of large, highly symmetric photonic devices. For computing symmetry-protected degenerate modes, our method reconstructs the modes numerically, offering a novel perspective for analyzing degenerate modes.

\section{Theory and methods}
\subsection{General framework}
\begin{figure*}[ht]
\centering
\includegraphics[width=17cm]{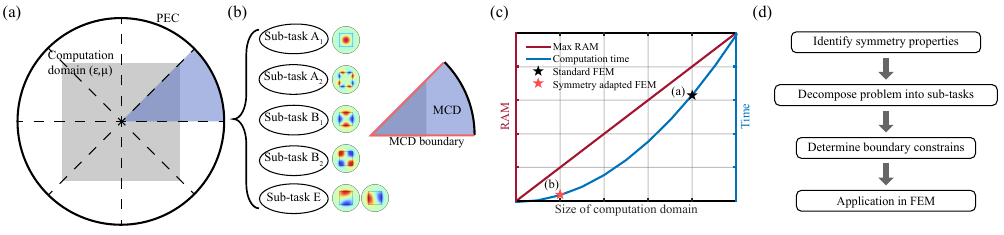}
\caption{Schematic of FEM modeling using combined symmetry. (a) Standard FEM for modal analysis on the $C_{4,v}$ structure. The entire domain satisfies the vector wave equation, with the boundary constraint being PEC. (b) CSA-FEM for $C_{4,v}$ structure. Only one-eighth of the entire domain (as MCD) is modeled, and it requires the calculation of five sub-tasks, each with different MCD boundary constraints. (c) RAM and computational time versus the size of computational domain. (d) General procedure of CSA-FEM.}
\label{fig1}
\end{figure*}
In modal analysis on photonic devices using standard FEM, it is necessary to determine the computational domain based on the device's structure and the boundary constraints at the domain edges. For instance, Fig. \ref{fig1}(a) displays a schematic diagram of the modal analysis for a $C_{4,v}$ structure, where PEC boundary condition is employed. The electric field $\boldsymbol{E}(\boldsymbol{r})$ of the modes within the computational domain satisfies the source-free vector wave equation,
\begin{equation}
\label{VectorWaveEq}
    \nabla \times \overline{\boldsymbol{\mu}}_{r}^{- 1}\nabla \times \boldsymbol{E}(\boldsymbol{r}) - k_{0}^{2}\overline{\boldsymbol{\epsilon}}_{r}\boldsymbol{E}(\boldsymbol{r})=0,
\end{equation} 
where $\nabla\times$ denotes the curl operator, $\overline{\boldsymbol{\mu}}_{r}$ 
($\overline{\boldsymbol{\epsilon}}_{r}$) is the relative permeability (permittivity), and $k_{0}$ is the vacuum wave number. In waveguide modal analysis with the propagation direction along the z-axis, the transverse and longitudinal components of electric field $\boldsymbol{E}(\boldsymbol{r})$  can be separated and written as $\boldsymbol{E}( \boldsymbol{r})=\left[\boldsymbol{E}_{t}({x,y}) + \hat{z}{E}_{z}({x,y})\right]e^{-\gamma z}$, where $\gamma$ is the complex propagation constant.

By utilizing combined symmetry, i.e. spatial symmetry within point groups and space-time symmetry (such as PT symmetry), we can model only a portion of the structure, referred to as the minimum computational domain (MCD), as shown in Fig. \ref{fig1}(b). The original problem corresponding to standard FEM is decomposed into multiple decoupled sub-tasks, each corresponding to a specific symmetry type of modes. Each sub-task need satisfy the boundary constraints related to the symmetry constraints on the MCD boundary. We refer to this method as combined symmetry adapted FEM (CSA-FEM). Due to the nonlinear relationship between computation time and the size of computation domain, although CSA-FEM increases the number of tasks to be solved, its computational efficiency—computation time and random access memory (RAM) usage—is significantly improved compared to standard FEM, as illustrated in Fig. \ref{fig1}(c). This improvement is attributed to the reduction in both the size of computational domain and the number of modes associated with each task, thereby decreasing the total computation time for the overall problem to the original problem. 

The implementation procedure of CSA-FEM is shown in Fig. \ref{fig1}(d), as divided into four steps,
\begin{enumerate}
    \item Identify the symmetry properties of the calculated structure, i.e., the associated symmetry group and the corresponding character table.
    \item Decompose the original problem into decoupled sub-tasks based on selected character table.
    \item Determine the symmetry constraints of each sub-task from the selected symmetry properties, and derive the boundary constraints that can be applied to FEM.
    \item Apply the corresponding boundary constraints to the sub-tasks and solve them to obtain the modes of different symmetry types. 
\end{enumerate}
For steps 1 and 2, the symmetry properties of the structure are generally evident. Once the symmetry group of the computational structure is identified, MCD and the number of sub-tasks can be determined, which is discussed in Subsection \ref{B2}. In this work, the symmetry group pertains to the point group, encompassing only spatial symmetry and excluding space-time symmetry in pseudo-Hermiticity systems. Space-time symmetry, i.e. PT symmetry, is not involved in sub-tasks decomposition, but significantly influences the size of MCD and MCD boundary constraints. The key to CSA-FEM lies in the step 3, which is discussed in Subsection \ref{B3}. In the combination of different symmetries, it is crucial to consider that the same symmetry imposes different constraints on different symmetry types of modes. Determining boundary constraints to ensure the solution of all modes is not straightforward and has never been examined, which will be studied shortly in a self-consistent manner with substantial details. Step 4 involves processing the matrix of the eigenvalue equation according to the boundary constraints when applying constraints in FEM, as discussed in Subsection \ref{B4}.

\subsection{Identify symmetry properties and decompose problem}\label{B2}
\begin{figure}[ht]
\centering
\includegraphics[width=8cm]{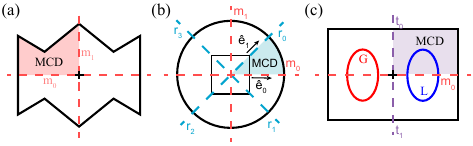}
\caption{Schematic examples of combined symmetry. (a) $C_{2,v}$ point group structure with mirror axes $m_{0}$ and $m_{1}$. (b) $C_{4,v}$ point group structure. (c) PT symmetry structure with $C_s$ point group.}
\label{fig2}
\end{figure}

In Figs. {\ref{fig2}}(a)-{\ref{fig2}}(c), we show three examples of combined symmetry, i.e., $C_{2,v}$, $C_{4,v}$ and the combination of PT symmetric and mirror symmetric structures respectively. In Fig. {\ref{fig2}}(a), the red dashed line $m_0/m_1$ represents mirror axis and the red domain indicates the MCD. Employing mirror symmetry allows for the complete restoration of the structure from MCD. The structure in Fig. {\ref{fig2}}(b) conforms to the $C_{4,v}$ point group, satisfying both mirror and rotational symmetries, thereby allowing the complete restoration of the structure from MCD using both mirror symmetry $\sigma_{v}$ and rotational symmetry $c_{4}$. The red dashed line $m_0$/$m_1$ also represents mirror axis, the blue dashed line $r_0$/$r_1$/$r_2$/$r_3$ represents axis of rotation and the blue domain indicates the MCD.The presence of additional symmetry shall further decrease MCD, and thereby enhances computational efficiency for CSA-FEM. As shown in Fig. {\ref{fig2}}(c), the structure is a combination of PT symmetry and mirror symmetry of $C_s$ point group, where the purple dashed line $t_0/t_1$ represents axis of PT symmetry. With the assistance of PT symmetry and mirror symmetry, one can restore the full solution of the complete structure from the purple MCD. 

\begin{table}[htbp]
\centering
\caption{\bf Character table for $C_{4,v}$ point group}
\begin{tabular}{cccccc}
\hline
 & $E$ & $2C_4$ & $C_2$  & $\sigma_{x}$ & $\sigma_{y}$\\
\hline
$A_{1}$ & $1$ & $1$ & $1$ & $1$ & $1$\\
$A_{2}$ & $1$ & $1$ & $1$ & $-1$ & $-1$\\
$B_{1}$ & $1$ & $-1$ & $1$ & $1$ & $-1$\\
$B_{2}$ & $1$ & $-1$ & $1$ & $-1$ & $1$\\
$E$     & $2$ & $0$ & $-2$ & $0$ & $0$\\
\hline
\end{tabular}
  \label{C4vtable}
\end{table}

Once the symmetry group that the optical structure satisfies is determined, the corresponding character table can be used to partition the sub-tasks. Using the $C_{4v}$ point group structure shown in Fig. \ref{fig2}(b) as an example, according to the character table, i.e., Table \ref{C4vtable}, the mirror symmetry and rotational symmetry categorize all modes into five different symmetry types, namely $A_1$, $A_2$, $B_1$, $B_2$ and $E$. The symmetry type of those modes is essentially the irreducible representation in formal group theory language.  Correspondingly, the original problem is also decomposed into five decoupled sub-tasks, each corresponding to a symmetry type of modes, as shown in Fig. \ref{fig1}(b). This decomposition provides inherent convenience for mode classification and analysis, enabling the characteristics of modes in different sub-tasks to be identified before computations are finalized\cite{bird_mode_2018}. The irreducible decomposition and the closure property of the symmetry group ensure the completeness and fidelity between the sub-tasks and the original problem, guaranteeing that all modes can be computed from all the sub-tasks. From a numerical perspective, CSA-FEM can automatically transform  the original large sparse matrix of standard FEM into block diagonal matrices, such that the computational efficiency can be significantly improved via parallel computing or other techniques (the efficiency in this work has been increased several times without the use of parallel techniques).

\subsection{Symmetry operation and boundary constraints}\label{B3}
The symmetry operations in group theory provide a theoretical basis to use combined symmetries in FEM modeling. Especially, the basic concepts on characters, character table and irreducible representation can be found in the textbook on group theory\cite{heine_group_2008,inui_group_2012}. The symmetry operation corresponding to symmetry $O$ is denoted as $P_O$. When a given optical structure exhibits symmetry $O$, all modes of this structure also possess the symmetry properties under the operation $P_O$. Specifically, acting $P_O$ on a mode $\boldsymbol{\psi}$ leads to the following relation,
\begin{equation}
\label{eqSymmetryOperation}P_{O}\boldsymbol{E}_{\boldsymbol{\psi}}(\boldsymbol{r})= \overline{M}\left(O\right)\boldsymbol{E}_{\boldsymbol{\psi}}(\boldsymbol{r}),
\end{equation}
where $\boldsymbol{E}_{\boldsymbol{\psi}}(\boldsymbol{r})$ is the electric field of mode $\boldsymbol{\psi}$, $\overline{M}\left(O\right)$ is the matrix of irreducible representation corresponding to symmetry $O$, which describes the constraints imposed by symmetry on modes of different sub-tasks. The character in the character table is the trace of $\overline{M}$. In the one-dimensional irreducible representation, the character and $\overline{M}$ are the same. For detailed information about symmetry operations, please refer to Supplement 1 or Xiong's work \cite{xiong_classification_2017}. Using Eq. ({\ref{eqSymmetryOperation}}), we can derive the boundary constraints of MCD based on the constraints of symmetry $O$. There are two situations outlined as follows,
\begin{enumerate}
\item When symmetry operation $P_O$ acts on non-degenerate modes $\psi_i$, $\overline{M}\left(O\right)$ degenerates into a scalar and is the character, and $\boldsymbol{E}_{\boldsymbol{\psi}}(\boldsymbol{r})=\boldsymbol{E}_{\psi_i}(\boldsymbol{r})$.
\item When symmetry operation $P_O$ acts on symmetry-protected degenerate modes $\psi_{1,2,\cdots,n}$, $\overline{M}\left(O\right)$ is a matrix and  $\boldsymbol{E}_{\boldsymbol{\psi}}(\boldsymbol{r})=\left[\boldsymbol{E}_{\psi_1}(\boldsymbol{r}),\boldsymbol{E}_{\psi_2}(\boldsymbol{r}),\cdots, \boldsymbol{E}_{\psi_n}(\boldsymbol{r})\right]^T$.
\end{enumerate}
In the second situation, in order to properly truncate the domain, we need to reconstruct a modal basis based on the principle of symmetry to conveniently apply constraints in FEM modeling. We use the $C_{4v}$ point group structure as an example to illustrate these two situations, as shown in Fig. \ref{fig2}(b). This structure encompasses both mirror and rotational symmetries.

Consider sub-task $A_1$ as an example of non-degenerate modes pertaining to the first situation. The $\overline{M}$ is selected corresponding to the symmetry type $A_1$, i.e., $\overline{M}(\sigma_{x})=1$, $\overline{M}(c_4)=1$. The spatial symmetry operation on the modes results in $P_O\boldsymbol{E}(\boldsymbol{r})=\overline{\boldsymbol{O}}\boldsymbol{E}(\overline{\boldsymbol{O}}^{-1}\boldsymbol{r})$, where $\overline{\boldsymbol{O}}$ is the matrix corresponding to symmetry $O$. Utilizing the mirror symmetry $\sigma_x$ and Eq. ({\ref{eqSymmetryOperation}}), we derive that $\overline{\boldsymbol{\sigma}}_{x}\boldsymbol{E}(\overline{\boldsymbol{\sigma}}_{x}^{-1}\boldsymbol{r})=\boldsymbol{E}(\boldsymbol{r})$, where $\overline{\boldsymbol{\sigma}}_{x}$ is the matrix corresponding to mirror symmetry $\sigma_x$. Hence, we derive the symmetry constraints on the MCD boundary as
\begin{subequations}
\begin{align}
\overline{\boldsymbol{\sigma}}_{x}\boldsymbol{E}(\boldsymbol{r}_{m_0})=\boldsymbol{E}(\boldsymbol{r}_{m_{0}}),\label{mirrorSY1}\\
\overline{\boldsymbol{\sigma}}_{x}\boldsymbol{E}(\boldsymbol{r}_{r_1})=\boldsymbol{E}(\boldsymbol{r}_{r_0}),\label{mirrorSY2}
\end{align}
\end{subequations}
where the MCD boundary $m_0$ and $r_0$ are represented by $\boldsymbol{r}=\boldsymbol{r}_{m_0}$ and $\boldsymbol{r}=\boldsymbol{r}_{r_0}$, respectively.
Similarly, the rotational symmetry $c_{4}$ leads to $\overline{\boldsymbol{c}}_{4}\boldsymbol{E}(\overline{\boldsymbol{c}}_{4}^{-1}\boldsymbol{r})=\boldsymbol{E}(\boldsymbol{r})$, where $\overline{\boldsymbol{c}}_{4}$ is the corresponding matrix. As $P_{c_4}$ acts on the MCD boundary $r_0$, one shall have,
\begin{equation}
\label{rotatSY1}    \overline{\boldsymbol{c}}_{4}\boldsymbol{E}(\boldsymbol{r}_{r_1})=\boldsymbol{E}(\boldsymbol{r}_{r_{0}}).
\end{equation}
Therefore, from Eq. (\ref{mirrorSY1}), we obtain $\hat{e}_{0}\times\boldsymbol{E}_{t}(\boldsymbol{r}_{m_0})=0$ and $E_{z}(\boldsymbol{r}_{m_0})=E_{z}(\boldsymbol{r}_{m_0})$. From Eqs. (\ref{mirrorSY2}) and (\ref{rotatSY1}), we obtain $\hat{e}_{1}\times\boldsymbol{E}_{t}(\boldsymbol{r}_{r_0})=0$ and $E_{z}(\boldsymbol{r}_{r_0})=E_{z}(\boldsymbol{r}_{r_0})$. This means that PMC is satisfied on both MCD boundary $m_0$ and $r_0$.

\begin{figure}[ht]
\centering
\includegraphics[width=8cm]{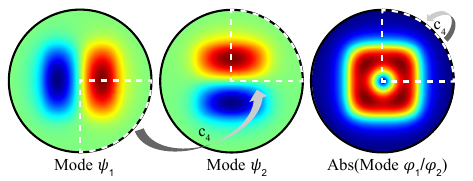}
\caption{Reconstruction of modes based on symmetry-protected degenerate modes. $c_4$ rotates $90^{\circ}$ counterclockwise. After rotating by $90^{\circ}$, $\phi_1$ and $\phi_2$ interchange, while $\varphi_1$ and $\varphi_2$ remain unchanged.}
\label{fig3}
\end{figure}

We proceed to discuss the second situation, wherein the two modes belong to sub-task $E$ and are doubly degenerate, i.e., $\boldsymbol{E}_{\boldsymbol{\psi}}=(\boldsymbol{E}_{\psi_1},\boldsymbol{E}_{\psi_2})^T$. Degenerate modes $\psi_1$ and $\psi_2$, shown in Fig. \ref{fig3}, are symmetric and anti-symmetric with respect to the mirror symmetry $\sigma_{x}$, respectively, as $P_{\sigma_{x}}\left( \begin{array}{c}\boldsymbol{E}_{\psi_1} \\ \boldsymbol{E}_{\psi_2}\end{array}\right)=\left(\begin{array}{cc} 1&0 \\ 0&-1\end{array}\right) \left( \begin{array}{c}\boldsymbol{E}_{\psi_1} \\ \boldsymbol{E}_{\psi_2}\end{array}\right)$. Rotational symmetry $c_{4}$ transforms degenerate modes into each other, that is $P_{c_{4}}\left( \begin{array}{c}\boldsymbol{E}_{\psi_1} \\ \boldsymbol{E}_{\psi_2}\end{array} \right)=\left(\begin{array}{cc} 0&1 \\ -1&0\end{array}\right) \left( \begin{array}{c}\boldsymbol{E}_{\psi_1} \\ \boldsymbol{E}_{\psi_2}\end{array}\right) $. According to the matrix diagonalization, we utilize $\psi_1$ and $\psi_2$ to reconstruct a pair of new degenerate modes, $\boldsymbol{E}_{\varphi_1}=\boldsymbol{E}_{\psi_1}+i\boldsymbol{E}_{\psi_2}$ and $\boldsymbol{E}_{\varphi_2}=\boldsymbol{E}_{\psi_1}-i\boldsymbol{E}_{\psi_2}$. The new modes $\varphi_1$ and $\varphi_2$ satisfy $P_{c_{4}}\left( \begin{array}{c}\boldsymbol{E}_{\varphi_1} \\ \boldsymbol{E}_{\varphi_2}\end{array}\right)=\left(\begin{array}{cc} i&0 \\ 0&-i\end{array}\right) \left( \begin{array}{c}\boldsymbol{E}_{\varphi_1} \\ \boldsymbol{E}_{\varphi_2}\end{array}\right)$ and $P_{\sigma_x}\left( \begin{array}{c}\boldsymbol{E}_{\varphi_1}\\ \boldsymbol{E}_{\varphi_2}\end{array}\right)=\left(\begin{array}{cc} 1&0 \\ 0&1\end{array}\right) \left( \begin{array}{c}\boldsymbol{E}_{\varphi_1}^{*} \\ \boldsymbol{E}_{\varphi_2}^{*}\end{array}\right)$. The new degenerate modes is completely decoupled under $\sigma_x$ and $c_4$, and one-eighth of the single mode field can be used to restore the complete mode field based on symmetry principle. It should be noted that it is actually impossible to distinguish between $\varphi_1$ and $\varphi_2$ in FEM, so the symmetry constraint corresponding to either one of them can be used. Hence, we derive the symmetry constraints on the MCD boundary for the degeneracy situation,
\begin{subequations}
\begin{align}
\overline{\boldsymbol{\sigma}}_{x}\boldsymbol{E}(\boldsymbol{r}_{m_0})&=\boldsymbol{E}^{*}(\boldsymbol{r}_{m_{0}}),\label{mirrorSY3}\\
\overline{\boldsymbol{\sigma}}_{x}\boldsymbol{E}(\boldsymbol{r}_{r_1})&=\boldsymbol{E}^{*}(\boldsymbol{r}_{r_0}),\label{mirrorSY4}\\
\overline{\boldsymbol{c}}_{4}\boldsymbol{E}(\boldsymbol{r}_{r_1})&=i\boldsymbol{E}(\boldsymbol{r}_{r_0}).\label{rotatSY2}
\end{align}
\end{subequations}
As a result, from Eq. (\ref{mirrorSY3}), we obtain $\hat{e}_{0}\times\boldsymbol{E}_{t,r}^{(E)}(\boldsymbol{r}_{m_0})=0$, $E_{z,r}^{(E)}(\boldsymbol{r}_{m_0})=E_{z,r}^{(E)}(\boldsymbol{r}_{m_0})$, $\hat{e}_{0}\cdot\boldsymbol{E}_{t,i}^{(E)}(\boldsymbol{r}_{m_0})=0$, and $E_{z,i}^{(E)}(\boldsymbol{r}_{m_0})=0$, where the subscript $r/i$ represents the real/imaginary part of the mode field. From Eqs.(\ref{mirrorSY4}) and (\ref{rotatSY2}), we obtain $\hat{e}_{1}\cdot\boldsymbol{E}_{t,i}^{(E)}(\boldsymbol{r}_{r_0})=\hat{e}_{1}\cdot\boldsymbol{E}_{t,r}^{(E)}(\boldsymbol{r}_{r_0})$ and $E_{z,i}^{(E)}(\boldsymbol{r}_{r_0})=E_{z,r}^{(E)}(\boldsymbol{r}_{r_0})$. The derivation of boundary constraints for the combination of spatial symmetry and space-time symmetry is identical to that for the combination of spatial symmetries.

\subsection{Finite element implementation for modal analysis}\label{B4}
In order to facilitate FEM to solve the eigenvalue problem, the variable substitution is used to make $\boldsymbol{e}_{t}=\gamma \boldsymbol{E}_{t}$, $e_{z}=E_{z}$. The transverse component $\boldsymbol{e}_{t}$ and the longitudinal component $e_{z}$ of the electric field are expanded on vector basis and scalar basis respectively as $\widetilde{\boldsymbol{e}}_t=\sum_{j=1}^{m}e_{t,j}\boldsymbol{\alpha}_{j}(x,y)$ and  $\widetilde{e}_z=\sum_{j=1}^{m}e_{z,j}{\alpha}_{j}(x,y)$, where $\widetilde{\boldsymbol{e}}_t$ and $\widetilde{e}_z$ denote the numerical electric field, $\boldsymbol{\alpha}_{j}(x,y)$ denotes the vector basis functions implemented by the first type of Nédélec elements, $e_{t,j}$ is the corresponding coefficient. It is worth noting that $\hat{e}\cdot \boldsymbol{\alpha}_{j}(x,y)=1$ and $\hat{e}\times \boldsymbol{\alpha}_{j}(x,y)=0$, thus $\hat{e}\cdot \boldsymbol{E}_{t}=0$ represents PEC and $\hat{e}\cdot \boldsymbol{E}_{t}=0$ represents PMC. ${\alpha}_{j}(x,y)$ denotes the scalar basis functions using Lagrange elements and $e_{z,j}$ is the corresponding coefficient. Implementing standard FEM procedure can lead to matrix form of Eq. (\ref{VectorWaveEq}),
\begin{equation}
\label{eq13}
    Ax=\lambda Bx,
\end{equation}
where $\lambda=\gamma ^{2}$ is the eigenvalue, A and B are system matrices, and $x$ is the DOFs vector.

In combination with rotational symmetry and mirror symmetry to calculate degenerate modes of sub-task $E$, the boundary constraints involve the real and imaginary parts of the mode fields. Meanwhile, Eq. (\ref{eq13}) can be written in the following form,
\begin{equation}
\label{eq14}
    \left( \begin{array}{cc}A&0\\0&A\end{array} \right)
    \left( \begin{array}{c}x_{r}\\x_{i}\end{array} \right)=
    \lambda \left( \begin{array}{cc}B&0\\0&B\end{array} \right)
    \left( \begin{array}{c}x_{r}\\x_{i}\end{array} \right),
\end{equation}
where $x_{r}$/$x_{i}$ is the real/imaginary part of $x$. From the boundary constraints in the preceding subsection, constraints on the DOFs can be derived as follows,
\begin{equation}
\label{eq16}
    x=Px',
\end{equation}
where $x=\left( \begin{array}{cccccc}x_{r,{MCD}} & x_{r,{r_0}} & x_{r,{m_0}} & x_{i,{MCD}} & x_{i,{r_0}} & x_{i,{m_0}} \end{array} \right)^{T}$ is the original DOFs vector, $P={\left( \begin{array}{cccccc} I&0&0&0&0&0\\0&I&0&0&I&0 \\0&0&I&0&0&0 \\0&0&0&I&0&0 \end{array} \right)^{T}}$ is the transformation matrix, $x'=\left( \begin{array}{cccc}x_{r,{MCD}}&x_{r,{r_{0}}}&x_{r,{m_0}}&x_{i,{MCD}} \end{array} \right)^{T}$ is the modified DOFs vector, the subscript $MCD$/$r_{0}$/$m_{0}$ represents the coefficients within the calculation domain/on the MCD boundary $r_{0}$/on the MCD boundary $m_{0}$, $I$ is the identity matrix. By using matrix operations and Eq. (\ref{eq16}), Eq. (\ref{eq14}) can be written as\cite{jin_finite_2015},
\begin{equation}
\label{eq17}
    P^{\dag}\left( \begin{array}{cc}A&0\\0&A\end{array} \right)Px'=
    \lambda P^{\dag} \left( \begin{array}{cc}B&0\\0&B\end{array} \right) Px'.
\end{equation}
So far, we have obtained the eigenvalue problem with imposed boundary constraints. Solving Eq. (\ref{eq17}) can result in the mode field and the effective refractive index.

When combining space-time symmetry and spatial symmetry, the boundary constraints also involve the real and imaginary parts of the mode fields. As it is a pseudo-Hermiticity system, Eq. (\ref{eq13}) can be written in the following form,
\begin{equation}
\label{eq18}
    \left( \begin{array}{cc}A_{r}&-A_{i}\\A_{i}&A_{r}\end{array} \right)
    \left( \begin{array}{c}x_{r}\\x_{i}\end{array} \right)=
    \lambda \left( \begin{array}{cc}B_{r}&-B_{i}\\B_{i}&B_{r}\end{array} \right)
    \left( \begin{array}{c}x_{r}\\x_{i}\end{array} \right).
\end{equation}
Similarly, boundary constraints imposed from combined PT symmetry and mirror symmetry can be included using identical procedures as described in Eqs. (\ref{eq16}) and (\ref{eq17}).

\section{Results and discussion} \label{Section3}
We validate the efficacy and efficiency of our method through two numerical examples, and compare the results obtained by CSA-FEM with those of standard FEM. Our method employs mesh density consistent with that of standard FEM and utilizes the same eigen solver. All numerical results are generated using MATLAB with home-made code. The MATLAB implementation of our CSA-FEM procedure is open-source and available on GitHub\cite{matlab-code}.
\subsection{PT-symmetric waveguide array}  \label{3A}
\begin{figure*}[ht]
\centering
\includegraphics[width=12.5cm]{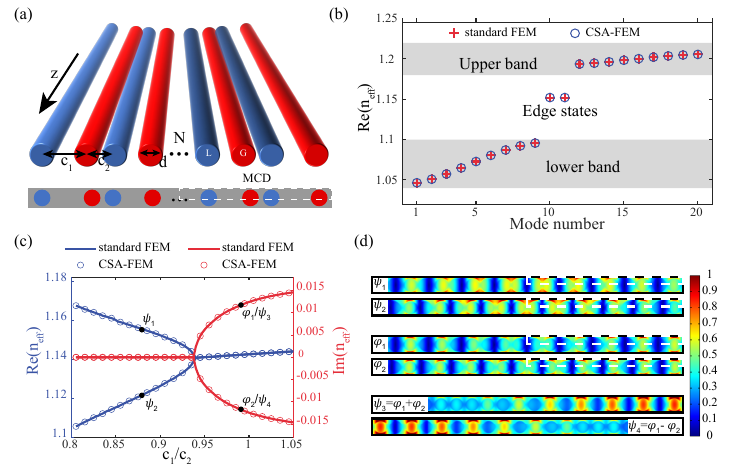}
\caption{pseudo-Hermiticity SSH model. (a) Schematics of the SSH model with N waveguides, where red and blue colors indicate gain and loss, respectively. (b) Band structure with $N=10$, $c_2=0.5\lambda_{0}$, $c_1/c_2=3$. (c) Exceptional point curve with N=10 by varying $c_1/c_2$. (d) Mode fields before and after PT symmetry breaking.}
\label{fig4}
\end{figure*}
To illustrate the combination of space-time symmetry and spatial symmetry, we use a pseudo-Hermiticity SSH model with PT symmetry defined by a finite coupled optical waveguide, as depicted in Fig. \ref{fig4}(a). The parameters $c_1$ and $c_2$ represent the distances between two adjacent waveguides, with red and blue colors symbolizing the gain and loss waveguides, respectively. $N$ is the number of waveguide pairs, $d$ is the diameter of the waveguides, and $\lambda_{0}$ is the vacuum wavelength. We set $N=10$, $c_2=0.5\lambda_{0}$, $d=0.4\lambda_{0}$,  $\epsilon_{r}=2.25+0.1i$ in the gain medium, and $\epsilon_{r}=2.25-0.1i$ in the loss medium. Given the mirror symmetry and PT symmetry of the structure, we can utilize CSA-FEM to enhance the efficiency of modal analysis, and the MCD is indicated by the white dashed line in Fig. \ref{fig4}(a). As shown in Fig. \ref{fig4}(b), at $c_1=3c_2$, the mode diagram of the SSH model displays a pair of edge states in the gap between the two bands. Red crosses represent standard FEM calculation results, while blue circles indicate those from CSA-FEM, both are consistent. 

Figure \ref{fig4}(c) illustrates the exceptional point curve for a pair of PT symmetry protected modes, where the straight lines represents calculation results from standard FEM  and the circles donate results obtained from CSA-FEM. Before PT symmetry breaking, from Eq. (\ref{eqSymmetryOperation}), we have $\boldsymbol{E}_{\psi_1/\psi_2}(\boldsymbol{r})=\boldsymbol{E}^{*}_{\psi_1/\psi_2}\left(-\boldsymbol{r}\right)$. After PT symmetry breaking, degenerate modes $\psi_3$/$\psi_4$ transforms into each other under PT symmetry. Similar to the treatment of degenerate modes in Subsection \ref{B3}, we reconstruct a new pair of modes, $\boldsymbol{E}_{\varphi_1}=\boldsymbol{E}_{\psi_3}-\boldsymbol{E}_{\psi_4}$ and  $\boldsymbol{E}_{\varphi_2}=\boldsymbol{E}_{\psi_3}+\boldsymbol{E}_{\psi_4}$, satisfying $P_{PT}\left( \begin{array}{c}\boldsymbol{E}_{\varphi_1} \\ \boldsymbol{E}_{\varphi_2}\end{array}\right)=\left(\begin{array}{cc} -1&0 \\ 0&1\end{array}\right) \left( \begin{array}{c}\boldsymbol{E}_{\varphi_1} \\ \boldsymbol{E}_{\varphi_2}\end{array}\right)$. Figure \ref{fig4}(d) shows the mode fields corresponding to modes $\psi_1$ and $\psi_2$ before PT symmetry breaking, restructured modes $\varphi_1$ and $\varphi_2$ after PT symmetry breaking, and recovered modes $\psi_3$ and $\psi_4$. These mode fields were all obtained through calculations using CSA-FEM and reconstructed over the entire domain by utilizing Eq. (\ref{eqSymmetryOperation}) from MCD. The relationship between $\psi_3$/$\psi_4$ and $\varphi_1$/$\varphi_2$ is essential for recovering the coupled mode fields after PT symmetry breaking. The recovery of modes $\psi_3$ and $\psi_4$ numerically proves that coupling occurred after PT symmetry breaking, and that this coupling can be decoupled through mode restructuring. When combining spatial symmetries, mode fields can similarly be recovered to the entire domain by applying Eq. (\ref{eqSymmetryOperation}). In Fig. \ref{fig4}(c), the computation time for the standard FEM is $72.5s$, whereas the total computation time for the CSA-FEM is only $19.7s$, representing approximately a fourfold improvement. This example demonstrates the effectiveness and efficiency of our method. 

\subsection{Hollow-core fiber} \label{3B}
\begin{figure*}[ht]
\centering
\includegraphics[width=15cm]{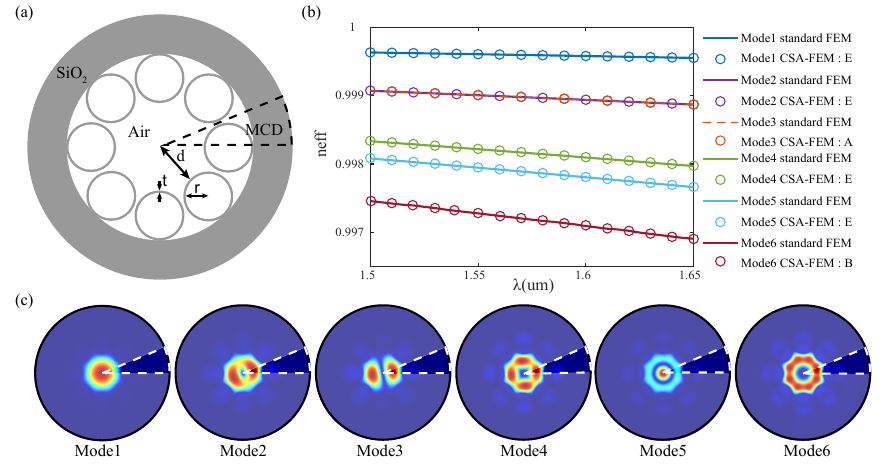}
\caption{Numerical example of a hollow-core fiber with the $C_{8,v}$ point group. (a) Structure of the hollow-core fiber, where MCD represents the CSA-FEM modelling domain. (b) Dispersion curves corresponding to six modes, where straight lines are results from standard FEM, and circles are results from CSA-FEM. (c) Mode fields of the six modes.}
\label{fig5}
\end{figure*}
The hollow regions of hollow-core fibers are typically much larger than the wavelength of light, leading to the support of multiple modes during the design phase. Additionally, the effective refractive indices of these various modes are almost identical, which further imposes challenges on the modal analysis. We perform modal analysis on a hollow-core fiber, as depicted in Fig. \ref{fig5}(a), where $d=20um$, $r=12um$, $t=0.5um$, $n_{SiO_{2}}=1.4378$, and the dashed black line represents the MCD. The hollow-core fiber satisfies the $C_{8,v}$ point group symmetries. Based on symmetry properties of rotational symmetry $c_8$ and mirror symmetry $\sigma_x$, one can divide the original problem into seven sub-tasks $A_{1}$, $A_{2}$, $B_{1}$, $B_{2}$, $E_{1}$, $E_{2}$, and $E_{3}$. We calculate the first 32 modes with higher effective refractive indices and plot the dispersion curves for six of these modes as shown in Fig. \ref{fig5}(b). The straight lines represent the results calculated by standard FEM, while the circles denote those calculated using CSA-FEM. Figure \ref{fig5}(b) clearly demonstrates that our method can separately calculate modes for different sub-tasks, effectively distinguishing them and significantly enhancing the analysis efficiency in the design of multimode fibers. Figure \ref{fig5}(c) displays the mode fields of the six selected modes. The white dashed line denotes the domain calculated by CSA-FEM, while the rest is depicted by the results from standard FEM, with an opacity set to $70\%$. The results obtained through our method are consistent with those from the standard FEM.

As shown in Table \ref{table2}, we summary the MCD, number of DOFs, and number of tasks in FEM modeling utilizing different symmetries. The number of modes calculated per task is approximately equal to the total number of modes divided by the number of tasks. Although CSA-FEM requires calculating seven sub-tasks separately, the total calculation time is only $123.5$ second, compared to $2861.6$ second for standard FEM, thereby enhancing the calculation speed by $23$ times. Therefore, CSA-FEM has more sub-tasks, smaller MCD, fewer DOFs, resulting in faster computational efficiency. When combining mirror symmetry and PT symmetry, although the DOFs remain the same as when using only mirror symmetry, the matrix changes from a complex matrix to a real matrix, effectively reducing the computational load by half. 
\begin{table*}[hb]
\centering
\begin{threeparttable}
\caption{\bf Summary of FEM symmetry, domain, number of DOFs and number of tasks}
\begin{tabular}{*5{c}}
\hline
 Symmetry & MCD & Number of DOFs & Number of tasks & Computation time\tnote{1}\\
\hline
- & $D$ & $M$ & $1$ & $2861.6$ second\\
$\sigma_x$ & $D/2$ & $M/2$ & $2$  & $1224.4$ second\\
$\sigma_x$, $\sigma_y$ & $D/4$ & $M/4$ & $4$  & $683.2$ second \\
$\sigma_x$, PT & $D/4$ & $M/2$ & $2$  & - \\
$\sigma_x$, $c_N$ & $D/2N$ & $M/N,M/2N$ & $\lfloor N/2 \rfloor$+\{2,3\}\tnote{2}  & $123.5$ second\\
\hline
\end{tabular}
    \begin{tablenotes}
        \item[1] Calculations were conducted using the structure shown in Fig. \ref{fig5}(a) as an example.
        \item[2] Taking the value of 3 if N is even, or 2 if N is odd.
    \end{tablenotes}
\label{table2}
\end{threeparttable}
\end{table*}
\section{Conclusion}
In summary, we systematically study the combination of different symmetries including spatial symmetry in point groups and space-time symmetry in pseudo-Hermiticity systems for numerical analysis of modal problem, and propose the CSA-FEM. Based on the irreducible representations of point groups and their enclosed property, we decompose the original problem into several decoupled sub-tasks, each corresponding to different symmetry type of modes and can recover the original problem. Compared to the standard FEM, our method not only enhances the efficiency of modal analysis significantly but also facilitates the design of complex optical waveguides by directly classifying modes based on their symmetry types. Furthermore, in dealing with symmetry-protected degenerate modes, we numerically reconstruct these modes, offering a novel perspective for understanding and computing degenerate modes. We provide the source code of CSA-FEM for download and use. Considering the fundamental role of symmetry in optics, this symmetry-based method is universal and easily extendable from FEM to other numerical methods such as FDM and transfer matrix method. We envisage that the general framework on symmetry adapted finite element modeling paves an important step towards symmetry adapted numerical modeling techniques beyond the optical finite element method.

\begin{backmatter}
\bmsection{Funding} 

\bmsection{Acknowledgments} 

\bmsection{Disclosures} 
The authors declare no conflicts of interest.

\bmsection{Data Availability Statement} 
Data underlying the results presented in this paper are not publicly available at this time but may be obtained from the authors upon reasonable request.

\bmsection{Supplemental document}
See Supplement 1 for supporting content

\end{backmatter}


\begin{thebibliography}{10}
\newcommand{\enquote}[1]{``#1''}

\bibitem{jin_topologically_2019}
J.~Jin, X.~Yin, L.~Ni, \emph{et~al.}, \enquote{Topologically enabled ultrahigh-{Q} guided resonances robust to out-of-plane scattering,} {\protect\JournalTitle{Nature}} \textbf{574}, 501--504 (2019). Publisher: Nature Publishing Group.

\bibitem{liu_circularly_2019}
W.~Liu, B.~Wang, Y.~Zhang, \emph{et~al.}, \enquote{Circularly {Polarized} {States} {Spawning} from {Bound} {States} in the {Continuum},} {\protect\JournalTitle{Physical Review Letters}} \textbf{123}, 116104 (2019). Publisher: American Physical Society.

\bibitem{dong_nanoscale_2022}
Z.~Dong, Z.~Mahfoud, R.~Paniagua-Domínguez, \emph{et~al.}, \enquote{Nanoscale mapping of optically inaccessible bound-states-in-the-continuum,} {\protect\JournalTitle{Light: Science \& Applications}} \textbf{11}, 20 (2022). Publisher: Nature Publishing Group.

\bibitem{wang_continuum_2023}
Q.~Wang, C.~Zhu, X.~Zheng, \emph{et~al.}, \enquote{Continuum of {Bound} {States} in a {Non}-{Hermitian} {Model},} {\protect\JournalTitle{Physical Review Letters}} \textbf{130}, 103602 (2023).

\bibitem{chang_multiple_2017}
M.-L. Chang, M.~Xiao, W.-J. Chen, and C.~T. Chan, \enquote{Multiple {Weyl} points and the sign change of their topological charges in woodpile photonic crystals,} {\protect\JournalTitle{Physical Review B}} \textbf{95}, 125136 (2017). Publisher: American Physical Society.

\bibitem{yang_ideal_2018}
B.~Yang, Q.~Guo, B.~Tremain, \emph{et~al.}, \enquote{Ideal {Weyl} points and helicoid surface states in artificial photonic crystal structures,} {\protect\JournalTitle{Science}} \textbf{359}, 1013--1016 (2018). Publisher: American Association for the Advancement of Science.

\bibitem{pan_real_2023}
Y.~Pan, C.~Cui, Q.~Chen, \emph{et~al.}, \enquote{Real higher-order {Weyl} photonic crystal,} {\protect\JournalTitle{Nature Communications}} \textbf{14}, 6636 (2023). Publisher: Nature Publishing Group.

\bibitem{khanikaev_photonic_2013}
A.~B. Khanikaev, S.~Hossein~Mousavi, W.-K. Tse, \emph{et~al.}, \enquote{Photonic topological insulators,} {\protect\JournalTitle{Nature Materials}} \textbf{12}, 233--239 (2013). Publisher: Nature Publishing Group.

\bibitem{song_detailed_2024}
S.~Song and L.~Zhang, \enquote{Detailed analysis of topological edge and corner states in valley-{Hall}-like photonic {Kagome} insulators,} {\protect\JournalTitle{Applied Physics B}} \textbf{130}, 41 (2024).

\bibitem{konotop_nonlinear_2016}
V.~V. Konotop, J.~Yang, and D.~A. Zezyulin, \enquote{Nonlinear waves in \${\textbackslash}mathcal\{{PT}\}\$-symmetric systems,} {\protect\JournalTitle{Reviews of Modern Physics}} \textbf{88}, 035002 (2016). Publisher: American Physical Society.

\bibitem{zhang_symmetry-breaking-induced_2019}
X.~Zhang, Q.-T. Cao, Z.~Wang, \emph{et~al.}, \enquote{Symmetry-breaking-induced nonlinear optics at a microcavity surface,} {\protect\JournalTitle{Nature Photonics}} \textbf{13}, 21--24 (2019). Number: 1 Publisher: Nature Publishing Group.

\bibitem{zhao_polarization-maintaining_2017}
J.~Zhao, M.~Tang, K.~Oh, \emph{et~al.}, \enquote{Polarization-maintaining few mode fiber composed of a central circular-hole and an elliptical-ring core,} {\protect\JournalTitle{Photonics Research}} \textbf{5}, 261 (2017).

\bibitem{liu_symmetrical_2018}
C.~Liu, W.~Su, Q.~Liu, \emph{et~al.}, \enquote{Symmetrical dual {D}-shape photonic crystal fibers for surface plasmon resonance sensing,} {\protect\JournalTitle{Optics Express}} \textbf{26}, 9039--9049 (2018). Publisher: Optica Publishing Group.

\bibitem{gao_design_2019}
J.~Gao, E.~Nazemosadat, C.~Yang, \emph{et~al.}, \enquote{Design, fabrication, and characterization of a highly nonlinear few-mode fiber,} {\protect\JournalTitle{Photonics Research}} \textbf{7}, 1354--1362 (2019). Publisher: Optica Publishing Group.

\bibitem{murphy_azimuthal_2023}
L.~R. Murphy and D.~Bird, \enquote{Azimuthal confinement: the missing ingredient in understanding confinement loss in antiresonant, hollow-core fibers,} {\protect\JournalTitle{Optica}} \textbf{10}, 854 (2023).

\bibitem{cong_symmetry-protected_2019}
L.~Cong and R.~Singh, \enquote{Symmetry-{Protected} {Dual} {Bound} {States} in the {Continuum} in {Metamaterials},} {\protect\JournalTitle{Advanced Optical Materials}} \textbf{7}, 1900383 (2019). \_eprint: https://onlinelibrary.wiley.com/doi/pdf/10.1002/adom.201900383.

\bibitem{droulias_chiral_2019}
S.~Droulias, I.~Katsantonis, M.~Kafesaki, \emph{et~al.}, \enquote{Chiral {Metamaterials} with \${PT}\$ {Symmetry} and {Beyond},} {\protect\JournalTitle{Physical Review Letters}} \textbf{122}, 213201 (2019). Publisher: American Physical Society.

\bibitem{feng_single-mode_2014}
L.~Feng, Z.~J. Wong, R.-M. Ma, \emph{et~al.}, \enquote{Single-mode laser by parity-time symmetry breaking,} {\protect\JournalTitle{Science}} \textbf{346}, 972--975 (2014). Publisher: American Association for the Advancement of Science.

\bibitem{kodigala_lasing_2017}
A.~Kodigala, T.~Lepetit, Q.~Gu, \emph{et~al.}, \enquote{Lasing action from photonic bound states in continuum,} {\protect\JournalTitle{Nature}} \textbf{541}, 196--199 (2017). Publisher: Nature Publishing Group.

\bibitem{hokmabadi_supersymmetric_2019}
M.~P. Hokmabadi, N.~S. Nye, R.~El-Ganainy, \emph{et~al.}, \enquote{Supersymmetric laser arrays,} {\protect\JournalTitle{Science}} \textbf{363}, 623--626 (2019). Publisher: American Association for the Advancement of Science.

\bibitem{chang_paritytime_2014}
L.~Chang, X.~Jiang, S.~Hua, \emph{et~al.}, \enquote{Parity–time symmetry and variable optical isolation in active–passive-coupled microresonators,} {\protect\JournalTitle{Nature Photonics}} \textbf{8}, 524--529 (2014). Publisher: Nature Publishing Group.

\bibitem{chen_parity-time-symmetric_2018}
W.~Chen, J.~Zhang, B.~Peng, \emph{et~al.}, \enquote{Parity-time-symmetric whispering-gallery mode nanoparticle sensor [{Invited}],} {\protect\JournalTitle{Photonics Research}} \textbf{6}, A23--A30 (2018). Publisher: Optica Publishing Group.

\bibitem{mortensen_fluctuations_2018}
N.~A. Mortensen, P.~a.~D. Gonçalves, M.~Khajavikhan, \emph{et~al.}, \enquote{Fluctuations and noise-limited sensing near the exceptional point of parity-time-symmetric resonator systems,} {\protect\JournalTitle{Optica}} \textbf{5}, 1342--1346 (2018). Publisher: Optica Publishing Group.

\bibitem{kim_parity-time_2024}
C.~Kim, X.~Lu, D.~Kong, \emph{et~al.}, \enquote{Parity-time symmetry enabled ultra-efficient nonlinear optical signal processing,} {\protect\JournalTitle{eLight}} \textbf{4}, 6 (2024).

\bibitem{nyakas_full-vectorial_2007}
P.~Nyakas, \enquote{Full-{Vectorial} {Three}-{Dimensional} {Finite} {Element} {Optical} {Simulation} of {Vertical}-{Cavity} {Surface}-{Emitting} {Lasers},} {\protect\JournalTitle{Journal of Lightwave Technology}} \textbf{25}, 2427--2434 (2007). Conference Name: Journal of Lightwave Technology.

\bibitem{zamani_aghaie_birefringence_2010}
K.~Zamani~Aghaie, S.~Fan, and M.~J.~F. Digonnet, \enquote{Birefringence {Analysis} of {Photonic}-{Bandgap} {Fibers} {Using} the {Hexagonal} {Yee}'s {Cell},} {\protect\JournalTitle{IEEE Journal of Quantum Electronics}} \textbf{46}, 920--930 (2010). Conference Name: IEEE Journal of Quantum Electronics.

\bibitem{otin_finite_2011}
R.~Otin, J.~Verpoorte, and H.~Schippers, \enquote{Finite {Element} {Model} for the {Computation} of the {Transfer} {Impedance} of {Cable} {Shields},} {\protect\JournalTitle{IEEE Transactions on Electromagnetic Compatibility}} \textbf{53}, 950--958 (2011). Conference Name: IEEE Transactions on Electromagnetic Compatibility.

\bibitem{andonegui_finite_2013}
I.~Andonegui and A.~J. Garcia-Adeva, \enquote{The finite element method applied to the study of two-dimensional photonic crystals and resonant cavities,} {\protect\JournalTitle{Optics Express}} \textbf{21}, 4072--4092 (2013). Publisher: Optica Publishing Group.

\bibitem{hiett_application_2002}
B.~P. Hiett, J.~M. Generowicz, S.~J. Cox, \emph{et~al.}, \enquote{Application of finite element methods to photonic crystal modelling,} {\protect\JournalTitle{IEE Proceedings - Science, Measurement and Technology}} \textbf{149}, 293--296 (2002). Publisher: IET Digital Library.

\bibitem{nicolet_modelling_2004}
A.~Nicolet, S.~Guenneau, C.~Geuzaine, and F.~Zolla, \enquote{Modelling of electromagnetic waves in periodic media with finite elements,} {\protect\JournalTitle{Journal of Computational and Applied Mathematics}} \textbf{168}, 321--329 (2004).

\bibitem{tavallaee_finite-element_2008}
A.~A. Tavallaee and J.~P. Webb, \enquote{Finite-{Element} {Modeling} of {Evanescent} {Modes} in the {Stopband} of {Periodic} {Structures},} {\protect\JournalTitle{IEEE Transactions on Magnetics}} \textbf{44}, 1358--1361 (2008). Conference Name: IEEE Transactions on Magnetics.

\bibitem{parisi_complex_2012}
G.~Parisi, P.~Zilio, and F.~Romanato, \enquote{Complex {Bloch}-modes calculation of plasmonic crystal slabs by means of finite elements method,} {\protect\JournalTitle{Optics Express}} \textbf{20}, 16690--16703 (2012). Publisher: Optica Publishing Group.

\bibitem{bagheriasl_bloch_2019}
M.~Bagheriasl, O.~Quevedo-Teruel, and G.~Valerio, \enquote{Bloch {Analysis} of {Artificial} {Lines} and {Surfaces} {Exhibiting} {Glide} {Symmetry},} {\protect\JournalTitle{IEEE Transactions on Microwave Theory and Techniques}} \textbf{67}, 2618--2628 (2019). Conference Name: IEEE Transactions on Microwave Theory and Techniques.

\bibitem{gao_hollow-core_2018}
S.-f. Gao, Y.-y. Wang, W.~Ding, \emph{et~al.}, \enquote{Hollow-core conjoined-tube negative-curvature fibre with ultralow loss,} {\protect\JournalTitle{Nature Communications}} \textbf{9}, 2828 (2018). Publisher: Nature Publishing Group.

\bibitem{roth_strong_2018}
P.~Roth, Y.~Chen, M.~C. Günendi, \emph{et~al.}, \enquote{Strong circular dichroism for the {HE}$_{\textrm{11}}$ mode in twisted single-ring hollow-core photonic crystal fiber,} {\protect\JournalTitle{Optica}} \textbf{5}, 1315--1321 (2018). Publisher: Optica Publishing Group.

\bibitem{kelly_gas-induced_2021}
T.~W. Kelly, P.~Horak, I.~A. Davidson, \emph{et~al.}, \enquote{Gas-induced differential refractive index enhanced guidance in hollow-core optical fibers,} {\protect\JournalTitle{Optica}} \textbf{8}, 916--920 (2021). Publisher: Optica Publishing Group.

\bibitem{mulvad_kilowatt-average-power_2022}
H.~C.~H. Mulvad, S.~Abokhamis~Mousavi, V.~Zuba, \emph{et~al.}, \enquote{Kilowatt-average-power single-mode laser light transmission over kilometre-scale hollow-core fibre,} {\protect\JournalTitle{Nature Photonics}} \textbf{16}, 448--453 (2022). Publisher: Nature Publishing Group.

\bibitem{mostafazadeh_pseudo-hermiticity_2002}
A.~Mostafazadeh, \enquote{Pseudo-{Hermiticity} versus {PT}-symmetry {III}: {Equivalence} of pseudo-{Hermiticity} and the presence of antilinear symmetries,} {\protect\JournalTitle{Journal of Mathematical Physics}} \textbf{43}, 3944--3951 (2002).

\bibitem{borgnia_non-hermitian_2020}
D.~S. Borgnia, A.~J. Kruchkov, and R.-J. Slager, \enquote{Non-{Hermitian} {Boundary} {Modes} and {Topology},} {\protect\JournalTitle{Physical Review Letters}} \textbf{124}, 056802 (2020).

\bibitem{stalhammar_classification_2021}
M.~Stålhammar and E.~J. Bergholtz, \enquote{Classification of exceptional nodal topologies protected by {PT} symmetry,} {\protect\JournalTitle{Physical Review B}} \textbf{104}, L201104 (2021).

\bibitem{bird_mode_2018}
D.~Bird, \enquote{Mode symmetry in microstructured fibres revisited,} {\protect\JournalTitle{Optics Express}} \textbf{26}, 31454 (2018).

\bibitem{heine_group_2008}
V.~Heine, \enquote{Group {Theory}: {Application} to the {Physics} of {Condensed} {Matter},} {\protect\JournalTitle{Physics Today}} \textbf{61}, 57--58 (2008).

\bibitem{inui_group_2012}
T.~Inui, Y.~Tanabe, and Y.~Onodera, \emph{Group theory and its applications in physics}, vol.~78 (Springer Science \& Business Media, 2012).

\bibitem{xiong_classification_2017}
Z.~Xiong, W.~Chen, P.~Wang, and Y.~Chen, \enquote{Classification of symmetry properties of waveguide modes in presence of gain/losses, anisotropy/bianisotropy, or continuous/discrete rotational symmetry,} {\protect\JournalTitle{Optics Express}} \textbf{25}, 29822--29834 (2017). Publisher: Optica Publishing Group.

\bibitem{jin_finite_2015}
J.-M. Jin, \emph{The finite element method in electromagnetics} (John Wiley \& Sons, 2015).

\bibitem{matlab-code}
J.~Wang, L.~Liu, Y.~Jing, \emph{et~al.}, \enquote{Hust-cpo/combined-symmetry-adapted-finite-element-modeling,}  (2024). \url{http://github.com/HUST-CPO/Combined-symmetry-adapted-finite-element-modeling/}.

\end{thebibliography}

\end{document}